\documentclass[aps,prb,twocolumn]{revtex4}
\usepackage{graphicx}
\def\bea{\begin{eqnarray}}
\def\eea{\end{eqnarray}}
\def\ben{\begin{equation}}
\def\een{\end{equation}}
\def\benu{\begin{enumerate}}
\def\enu{\end{enumerate}}
\def\la{{\langle\, }}
\def\ra{{\,\rangle }}
\def\pm{\prime}
\def\cd{c^{\dagger}}
\def\pl{\parallel}

\def\dos{{\mathcal N}}
\def\ek{\epsilon_{k}}
\def\e {\epsilon} 
\def\om{\omega}
\def\a{\alpha}
\def\b{\beta} 
\def\r{\rho}  
\def\s{\sigma}
\def\t{\tau} 
\def\d{\delta}
\def\th{\theta}
 
\def\g{\gamma}

\def\bk{{\bf k}}
\def\bq{{\bf q}}
\def\bp{{\bf p}}
\def\bQ{{\bf Q}}
\def\bS{{\bf S}}
\def\bP{{\bf \Pi}}
\def\bj{{\bf j}}
\def\ham{{\mathcal H}}

\begin{document}

\title{Thermoelectric Behaviour Near Magnetic Quantum Critical Point}

\author{Indranil Paul and Gabriel Kotliar}

\affiliation{Center for Materials Theory, Department of Physics and Astronomy,
Rutgers University, Piscataway, New Jersey 08854}
\date{\today}

\begin{abstract}
We use the  coupled  2d-spin-3d-fermion model  proposed
by Rosch {\sl et. al.} (Phys. Rev. Lett.  {\bf 79}, 159 (1997))
to study the thermoelectric behaviour 
of a heavy fermion compound when it is close to an antiferromagnetic 
quantum critical point. When the low energy spin fluctuations are quasi two
dimensional, as has been observed in ${\rm YbRh}
_2{\rm Si}_2$ and $ {\rm CeCu}_{6-x}{\rm Au}_x $,  with a typical  
2d ordering wavevector  and 3d  Fermi surface, the ``hot''  regions
on the Fermi surface have  a finite area. 
Due to enhanced scattering with the nearly critical spin 
fluctuations, the electrons in the hot region are strongly renormalized. We 
argue that there is an intermediate energy scale where the  qualitative 
aspects of the renormalized hot electrons are captured by a weak-coupling 
perturbative calculation. Our examination of the electron self energy shows 
that the entropy carried by the hot electrons is larger than usual. This 
accounts for the anomalous logarithmic temperature dependence of specific heat
observed in these materials. We show that the same mechanism 
produces logarithmic temperature dependence in thermopower. This has been 
observed in $ {\rm CeCu}_{6-x}{\rm Au}_x $. We expect to see the same 
behaviour from future experiments on ${\rm YbRh}_2{\rm Si}_2$.  

\end{abstract}

\maketitle

\section{Introduction}

Understanding the behaviour of a system close to antiferromagnetic quantum 
critical point (QCP) is currently an area of active research. The problem is 
interesting both in the context of high temperature superconductors as well as
heavy-fermion materials, especially to understand metallic phases that show
non-Fermi liquid (NFL) properties. In recent times several materials have been
discovered where it has been possible to demonstrate the existence of magnetic
QCP.~\cite{loh,trov,mathur} This has made the problem an exciting ground 
where theoretical  
understanding of electrons with strong correlations can be verified 
experimentally. One  central issue in this problem is an appropriate 
theoretical treatment
of electrons interacting with spin fluctuations close to the QCP
where magnetic correlation length diverges. 
A second central issue, is whether the spin fermion model describes the
relevant degrees of freedom, or whether a more basic model, allowing for the
disintegration of the  binding of  local moments to the quasiparticles,
is necessary for describing this transition.\cite{schr00,si}

In this paper we will discuss two experimentally well-studied heavy fermion
materials,  $ {\rm CeCu}_{6-x}{\rm Au}_x $~\cite{loh} and 
${\rm YbRh}_2{\rm Si}_2$,~\cite{trov}
that exhibit antiferromagnetic QCP. In doped ${\rm CeCu}_6$, replacing Cu with
larger Au atoms, favours the formation of long range magnetic order.~\cite
{loh} Beyond a 
critical doping $x_c = 0.1$, the ground state of the system is 
antiferromagnetic with finite N\'{e}el temperature ($T_N$).~\cite{ros97}
At the critical doping
$T_N$ is zero and the system has a QCP. On the other hand ${\rm YbRh}_2{\rm Si
}_2$ is undoped and atomically well-ordered.~\cite{trov} It is a much cleaner 
material 
than $ {\rm CeCu}_{6-x}{\rm Au}_x $, with residual resistivity ($\r_0$) 
smaller by a factor of about 10.
At ambient pressure it develops long range magnetic
order at a very low temperature of $T_N \simeq 65$ mK.~\cite{trov} The 
ordering 
temperature can be suppressed to practically zero (less than $20$ mK) by 
applying a magnetic field of only $45$ mT.~\cite{trov} Both these materials 
show 
pronounced deviations from Fermi liquid (FL) behaviour, which is believed to
be due to closeness to the QCP. For instance, the dependence of electrical
resistivity $\Delta \r = \r - \r_0$ to temperature $T$ is $\Delta \r \propto
T$, while that of specific heat $C$ is $C/T \propto - \ln T$.~\cite{loh,trov} 
This is in 
contrast with FL behaviour which predicts $\Delta \r \propto T^2$ and 
$C/T = {\rm constant}$. The low temperature NFL behaviour is 
observed over a decade of temperature, up to about $1$ K for 
$ {\rm CeCu}_{6-x}{\rm Au}_x $,~\cite{loh,ros97} and up to as high as $10$ K 
for ${\rm YbRh}_2{\rm Si}_2$.~\cite{trov}
The source of the interesting physics in these materials is the localized
$4f$ electrons~\cite{schr00} of Ce$^{3+}$ (in $4f^1$ electronic configuration)
and Yb$^{3+}$
(in the configuration $4f^{13}$), and their interaction with the relatively
delocalized $s$, $p$ and $d$ orbital electrons that form a conduction band 
with a well defined Fermi surface at low temperature. The conduction 
electrons and the localized $4f$
electrons carrying magnetic moment are coupled by exchange interaction ($J$).
Below a certain critical value of exchange interaction ($J_c$), the local 
moments interact with each other,
mediated by conduction electrons, and at sufficiently low temperature form 
long range antiferromagnetic order. On the other hand, if the exchange 
coupling is strong ($J > J_c$), the local moments are quenched below a 
certain temperature
(lattice Kondo temperature). The quenched moments hybridize with the 
conduction electrons and they participate in the formation of the Fermi sea.
The ground state of such a system is non-magnetic. The exchange coupling is 
usually tuned experimentally by either doping the material or by applying 
external pressure or external magnetic field.

For $ {\rm CeCu}_{6-x}{\rm Au}_x $ there are two different 
views~\cite{schr00} regarding the nature of the system in the non-ordered 
phase and the 
corresponding mechanism by which the critical instability occurs. In the first
picture, the lattice Kondo temperature ($T_K^*$) becomes zero exactly at the 
critical point ($ J= J_c$). The local moments of the $4f$ electrons survive 
at all finite temperature close to
the critical point. At the transition point they are critically quenched. The
local moments produce the critical magnetic fluctuations that destabilize the
Fermi sea. It has been argued, in favour of this mechanism, that the data on 
magnetic susceptibility shows non-trivial scaling 
with temperature.~\cite{schr98} At the critical point the susceptibility has 
the scaling
form $\chi = T^{- \a}f(\om/T)$ with an anomalous exponent $\a \simeq 0.75$,
which is different from conventional insulating magnets which have $\a = 1$. 
The alternative picture suggests that $T_K^*$ is finite at the 
critical point. Well below this temperature, and close to the 
critical point, the local moments are quenched by Kondo mechanism. The
$4f$ electrons become part of the Fermi sea. Then, the phase transition 
occurs by the usual spin-density wave instability of the Fermi surface. In 
this picture the local moments do not play any role in the phase transition.
This  theoretical viewpoint, proposed  by Rosch and collaborators,~\cite
{ros97} is motivated by  inelastic neutron
scattering data on $ {\rm CeCu}_{5.9}{\rm Au}_{0.1} $ which show that the 
nearly critical spin fluctuations are two dimensional.~\cite{stock} 
But the  origin of the 
quasi 2d behaviour of spin fluctuations is not well understood. However, the 
same feature is probably also present in ${\rm YbRh}_2{\rm Si}_2$, where the 
structure of the lattice provides a more natural explanation for the spin 
fluctuations to be 2d.~\cite{trov} Besides the nature of the magnetic 
correlations, there 
are different opinions regarding the dynamics of the spin fluctuations. It 
has been argued~\cite{subir95} that if the ordering wave-vector spans 
different points of 
the Fermi surface, then the dynamics of the spin fluctuations is overdamped, 
with dynamic exponent $z=2$. This model of spin fluctuations 
with $d=2$ and 
$z=2$, coupled with three-dimensional electrons, was used to explain the
linearity of resistivity with temperature.~\cite{ros97} Following the method 
of Hertz,~\cite{hertz,millis93} in 
which the system is described entirely in terms of the spin fluctuations, 
after a formal Hubbard-Stratonovich transformation to integrate the fermion 
modes, it also explains the logarithmic 
temperature dependence of specific heat.~\cite{ros97,millis93} In an 
alternative description,~\cite{chubu1} in 
terms of low energy electrons interacting with spin fluctuations, it has been 
suggested recently that both the frequency and momentum dependence of the 
spin fluctuation propagator undergo singular corrections such that the 
propagator acquires an anomalous dimension $\eta \sim 1/4$.~\cite{chubu2}
Thus, after 
nearly a decade, there is still no clear understanding regarding the 
appropriate model that describes the quantum phase transition. 

In this paper we will study the thermoelectric behaviour of a system in the 
paramagnetic phase and close to antiferromagnetic QCP. For $ {\rm CeCu}_{5.9}
{\rm Au}_{0.1} $ it is known that thermopower ($S_t)$ has a dependence similar
to specific heat over the same range of temperature,~\cite{benz,pflei} i.e, 
$S_t/T \propto - \ln T$.
We will show that scattering with nearly critical spin fluctuations give rise
to temperature dependent quasiparticle mass ($m^*$) over much of the Fermi 
surface. The signature of this can be seen in static response (specific heat)
and in transport (thermopower). Finally we will argue that the same mechanism
should be relevant for ${\rm YbRh}_2{\rm Si}_2$, and so we expect to see the 
same behaviour for thermopower from future experiments.

\vspace*{0.18cm}

\section{Model}

Our model is motivated by the second picture as described above. It assumes
that $T_K^*$ defines a high energy parameter. For $T \sim T_K^*$ the local 
nature of the spins of the $4f$ electrons is important as they participate in
some lattice Kondo phenomenon. For $T < T_K^*$, the $4f$ electrons become part
of the hybridized conduction band. In this regime the nearly critical spin
fluctuations of the conduction electrons is important.
It is an intermediate 
temperature range where the system is described by low energy conduction 
electrons interacting with quasi 2d spin fluctuations.
Within the spin-fermion description,  at sufficiently low temperature,  
the 3d nature of the spin fluctuations is retrieved and 
the model used here  ceases to  be valid. 
In this regime, the model predicts, in pure systems, a crossover
to an electronic Fermi liquid with a finite mass. However the
physics governing this dimensional crossover, has not been investigated.

The model is described by the Hamiltonian
\begin{widetext}
\bea
\ham &=& \sum_{\bk,\s} \ek \cd_{\bk,\s} c_{\bk,\s} + \frac{g_0}{2} \sum_{\bk,
\bq,\a,\b} \cd_{\bk + \bq,\a} c_{\bk,\b} {\bf \s}_{\a,\b} \cdot \bS_{-\bq}
+ \sum_{\bq} \left[ \chi^{-1}(\bq) \bS_{\bq}\cdot \bS_{-\bq}
+ \bP_{\bq} \cdot \bP_{-\bq} \right] 
\nonumber \\ && 
+ \frac{u_0}{4} \sum_{\bk_1,\ldots ,\bk_4} \left( \bS_{\bk_1}\cdot 
\bS_{\bk_2} \right) \left( \bS_{\bk_3}\cdot \bS_{\bk_4} \right)
\d(\bk_1+ \ldots + \bk_4). 
\eea
\end{widetext}
Here $\cd_{\bk,\s}$ is the electron creation operator, $\bS_{\bq}$ is the 
operator for the spin fluctuations, $\bP_{\bq} = \partial_t 
\bS_{\bq}$ is the conjugate momentum field for the spin fluctuations, and 
$\chi(\bq)$ is the static magnetic susceptibility. $g_0$ is the bare coupling 
between the electrons and the spin fluctuations, and $u_0$ is the interaction
energy of the spin fluctuations. The collective spin fluctuations are formally
obtained by integrating out high energy electrons in the band up to a certain 
cutoff.~\cite{chubu1} Thus the typical energies of the spin fluctuations 
$\om_s \sim W$, 
the bandwidth of the conduction electrons. The system is close to an 
antiferromagnetic instability with ordering wave-vector $\bQ$. We will assume
that the dynamics of the spin fluctuations is purely damped with 
dynamic exponent $z=2$. The spectrum of the 2d spin fluctuations will be 
described by~\cite{hertz,millis93}
\ben
\chi^{-1}(\bq, \om) = \d + \om_s (\bq - \bQ)^2_{\pl} -i\g \left| \om \right|.
\een
Here $\d$ is the mass of the spin fluctuations and measures the deviation from
the QCP, the parallel directions are those along the planes of magnetic 
correlation, and $\g \sim (g_0/\e_F)^2$ is an estimate of the damping from 
the polarization bubble. In the spin fluctuation part of the Hamiltonian, the 
interaction term $u_0$ is marginal, since the scaling dimension is zero.~\cite
{hertz,millis93} The main contribution of this term is to renormalize the mass
of the spin fluctuations ($\d$) and make it temperature dependent. Within a 
Gaussian approximation, $\d$ is linearly dependent on temperature, up to 
logarithmic corrections.~\cite{ros97,millis93}
We will ignore other effects of the $u_0$ term in our discussion, and will 
consider only the quadratic term with temperature dependent mass of the spin 
fluctuations.
To simplify the calculation we will assume a spherical
Fermi surface for the non-interacting electrons, with the ordering wave-vector
$\bQ= (\a,0,2k_F \cos \th_0)$. Here $\th_0 \neq 0$ (i.e. not $2k_F$ ordering),
and $\th_0 \neq \pi/2$ (i.e. not ferromagnetic ordering). We have chosen 
${\bf\hat{x}}$ as the direction along which the spin fluctuations are 
uncorrelated, and $\a$, the ordering in the $x$-direction, varies from one 
plane of magnetic correlation to another. Since the spectrum of spin 
fluctuations is 2d, those carrying momentum of the form $\bQ + a{\bf\hat{x}}$,
where $a$ is arbitrary, are all nearly critical. Due to constraints from 
energy-momentum conservation, only those points on the Fermi surface that are 
connected by the nearly critical spin fluctuations are particularly sensitive
to the QCP, since electrons at these points undergo singular scattering with 
the spin fluctuations. These are the so-called ``hot spots''. It is important
to note that since the spin fluctuations are 2d, 
there will be a finite area of the Fermi surface that is hot. Though it is 
worthwhile to estimate the fraction of the Fermi surface that is hot, 
theoretically it is a daunting task. In our calculation we will assume that 
most of the Fermi surface is hot. In effect, we are assuming that contribution
to static response and also to transport is mostly from the hot regions. It 
was pointed out by Hlubina and Rice~\cite{rice} that in transport the hot 
carriers are 
less effective than the cold ones. This is because the quasiparticle lifetime 
of the hot carriers is less than that of the cold carriers, since the former
suffer enhanced scattering with the spin fluctuations. As we will show below, 
the lifetime of the hot electrons $\t_h \propto 1/T$, while the cold electrons
have Fermi liquid characteristics with $\t_c \propto 1/T^2$. If $x$ is the 
fraction of the Fermi surface (FS) that is hot, then we can make an estimate 
of conductivity $\s$, 
\[
\s \propto \la \t_{\bk} \ra_{{\rm FS}} \propto \frac{x}{T/\e_F} + \frac{1-x}
{(T/\e_F)^2}.
\]
The first term, which is the contribution from the hot region, will dominate 
to give $\Delta \r \propto T$ only if $x> 1/(1+ T/\e_F)$. This gives a 
rough estimate of the fraction necessary for the hot carriers to dominate.
In the case of $ {\rm CeCu}_{6-x}{\rm Au}_x $, which is a dirtier material,
the above estimation is more involved. It was recently shown~\cite{ros99} 
that the effect 
of disorder is to favour isotropic scattering and thereby reduce the 
effectiveness of Hlubina-Rice mechanism. Thus, one should expect a smaller 
fraction, than estimated above, enough to make the contribution of the hot 
carriers significant for $ {\rm CeCu}_{6-x}{\rm Au}_x $.
  
\section{Electron Self Energy}
 
To calculate the effect of the low energy spin fluctuations on the hot 
electrons, we will examine the electron self-energy. The lowest order term
in perturbation gives,
\ben
\Sigma \left(\bp, \om \right) = - \frac{g_0^2}{V}  \sum_{\bk} \int_{-\infty}
^{\infty} \frac{d \Omega}{2\pi i} \ \chi \left(\bk, \Omega \right) 
G \left( \bp + \bk, \omega + \Omega \right), 
\een
where $G\left(\bp, \om \right)$ is the free electron propagator given by,
\[
G\left(\bp, \om \right) = \frac{n_p}{\om - \e_p -i \eta} +
\frac{1- n_p}{\om - \e_p +i \eta}.
\]
Here $ n_p $ is the electron occupation of the momentum state $\bp$ at $T=0$.
As expected, the above expression has different behaviours in the hot and 
cold regions. But within each region the self-energy is practically momentum 
independent. 
The imaginary part of the self-energy gives the quasiparticle lifetime as 
determined by scattering with the spin fluctuations. For $\om >0$ we have,
\bea
\lefteqn{
{\rm Im}\Sigma \left( \bp, \om \right) =} 
\nonumber \\ 
&& - \frac{g_0^2}{V} \sum_{0 < \ek<\om}
\frac{\left( \om - \e_{\bk +\bp} \right)}{\left( \d + \om_s \left(
\bk -\bQ \right)^2_{\pl} \right)^2+\left(\om - \e_{\bk +\bp} \right)^2}.
\nonumber
\eea
If $\bp$ is a point in the hot region, then it is connected to 
another hot spot by a wave-vector of the form $\bk = \bQ + a{\bf\hat{x}}$. We 
linearize the spectrum about this second hot point, and perform the integral 
in terms of local coordinates around it. In the hot region we get,
\ben
{\rm Im}\Sigma \left( \bp, \om \right) \propto - \left( \frac{g_0^2}{\e_F 
\om_s} \right) \frac{\om^2}{{\rm max}[\d,\om]}.
\een
For $\om > \d$ the lifetime of the hot electrons is much smaller than 
that given by Fermi
liquid behaviour (${\rm Im}\Sigma(\om) \propto \om^2$). As we have mentioned 
above, this is due to more effective scattering with the spin fluctuations in 
this region. For the cold electrons the behaviour is Fermi liquid like. 

Next, 
we will examine the real part of the self-energy. 
The dependence of ${\rm Re} \Sigma$ on frequency is more important than the 
dependence on momentum. We get, 
\begin{widetext}
\ben
- \lim_{\om \rightarrow 0} \frac{\partial}{\partial \om}{\rm Re} \Sigma 
\left( \bp, \om \right)=
\frac{g^2_0}{\pi V}\sum_{\bk} \left\{ \frac{1}{\g_{\bk}^2 + \e_{\bk+\bp}^2}
- \frac{\left( \g_{\bk}^2 - \e_{\bk+\bp}^2 \right)}{\left(
\g_{\bk}^2 + \e_{\bk+\bp}^2 \right)^2} \ln \left| \frac{\g_{\bk}}
{\e_{\bk+\bp}} \right| + \frac{\pi (2n_{\bk+\bp}-1) \g_{\bk}\e_{\bk+\bp}}
{\left(\g_{\bk}^2 + \e_{\bk+\bp}^2 \right)^2} \right\}.
\een
\end{widetext}
Here $\g_{\bk}= \d + \om_s(\bk -\bQ)^2_{\pl}$. If $\bp$ is a point 
within the hot region, each of the three terms in the above expression is 
logarithmic. As before, after linearizing the spectrum near the second hot 
spot, we get,
\ben
- \lim_{\om \rightarrow 0} \frac{\partial}{\partial \om}{\rm Re} \Sigma 
\left( \bp, \om \right) \propto \left( \frac{g_0^2}{\pi \e_F \om_s} \right)
\ln \left( \frac{\om_s}{\d} \right).
\een
Due to scattering, the non-interacting electron mass $m$ is renormalized to
the quasiparticle mass $m^* = m/Z$ (in the absence of any momentum dependence
of the electron self-energy), where
\[
Z^{-1} = 1 - \lim_{\om \rightarrow 0} \frac{\partial}{\partial \om}{\rm Re} 
\Sigma \left( \bp, \om \right)
\]
defines the quasiparticle residue.
Since $\d$, which measures the deviation from the critical point, can be 
written as $\d = \Gamma ( p - p_c) + T$, the quasiparticle mass becomes 
temperature dependent. Here $p$ is an experimental parameter that can be tuned
to the critical value $p_c$, and $\Gamma$ is an appropriate energy parameter.
As a consequence the entropy of each hot quasiparticle becomes 
anomalously large. This can be seen from the expression for entropy ($S$) per
particle,~\cite{agd}
\[
\frac{S}{N} = \sum_{\bp} \frac{1}{\pi T} \int_{- \infty}^{\infty} d \om
\left(- \frac{\partial f}{\partial \om} \right) \om \ tan^{-1} \left( 
\frac{\t(\om)}{\e_p - \om/Z} \right). 
\]
Here $f(\om)$ is the Fermi function, and $\t(\om)$ is quasiparticle lifetime
obtained from the inverse of imaginary part of self energy. From the above 
expression it is easy to 
see that $S/N \propto 1/Z$. Over the hot region, keeping only the leading 
term, $Z^{-1} \sim \ln(1/\d)$. Then,
\ben
S/N  \propto \dos (0) T \left( \frac{g^2_0}{\e_F \om_s} \right) \ln 
\left( \frac{\om_s}{\d} \right),
\een
where $\dos (0)$ is the density of states of the non-interacting system at 
the Fermi energy. For $T > \Gamma( p - p_c)$, the temperature dependence of 
entropy is $S \propto T \ln (1/T)$, which is different from Fermi liquid 
behaviour ($S \propto T)$. This gives rise to the anomalous logarithmic 
temperature dependence
of specific heat. In the past~\cite{millis93,ros97} this behaviour has 
been understood from a purely
bosonic point of view following the formalism of Hertz and Millis. For the 
spin fluctuations the Gaussian part of the action gives a free energy $F 
\propto T^2 \ln T$, which explains the $\ln (1/T)$ behaviour of $C/T$. 
Thus, here we find that there is agreement between the results of the 
spin-fermion model and the pure bosonic model.

\section{Thermopower}

From our discussion on entropy, it is natural to expect that this 
entropy enhancement should be seen in the measurement of thermopower ($S_t$). 
This is because one can think of thermopower as proportional to the 
correlation function between the
heat current and the particle current, and heat current involves the 
transport of entropy
due to temperature and electric potential gradients in the system. 
Strictly speaking, thermopower is defined as a ratio of two correlation 
functions,~\cite{mahan} i.e,
\[
S_t = \frac{L_{12}}{e T L_{11}},
\]
where
\[
L_{12} = \lim_{\om \rightarrow 0} \frac{1}{\om V}{\rm Im}\int_{0}^{\beta}
d\t e^{i\om\t}\la T_{\t} \bj_Q(\t)\cdot \bj(0) \ra,
\]
is the correlation function between heat current ($\bj_Q$) and particle 
current ($\bj$), and 
\[
L_{11} = \lim_{\om \rightarrow 0} \frac{1}{\om V}{\rm Im}\int_{0}^{\beta}
d\t e^{i\om\t}\la T_{\t} \bj(\t)\cdot \bj(0) \ra, 
\]
is the correlation function between particle currents. $L_{11}$ is a measure 
of 
electrical conductivity ($\s = e^2 L_{11}$). Here we are ignoring the tensor
nature of $L_{11}$ and $L_{12}$, and assuming that temperature and 
potential gradients and the thermal current are along the major symmetry 
directions of the lattice so that the tensors are diagonal. 
We express the single particle energies with respect to the chemical potential
and assume that chemical potential in the sample is uniform. The 
expression for heat current is given by,
\[
\bj_Q = \frac{i}{2} \sum_{\bp,\s} {\bf v}_{\bp} \left( \cd_{\bp,\s} \dot{c}
_{\bp,\s} - \dot{\cd}_{\bp,\s} c_{\bp,\s} \right).
\]
In principle, heat current will have a second term of the form 
$\frac{i}{2} \sum_{\bk,\s} {\bf \nabla}_{k} U(\bk) ( \dot{n}_{\bk,\s} 
n_{-\bk,\s} - \dot{n}_{-\bk,\s} n_{\bk,\s} )$, where $U(\bk)$ is the Fourier
transform of the interaction term between the electrons. However, such a term 
is quartic in fermionic operators and generates only subleading contributions
in our calculation. We will also ignore corrections to the particle current 
and heat current vertices due to exchange of spin fluctuations. These vertex
corrections are nonsingular, and change only the numerical prefactor (which
we do not attempt to calculate) of our leading term,  because the spin 
fluctuations are peaked around a finite wave-vector. With these approximations
the expressions for the correlation functions can be re-expressed in a more 
transparent form as
\[
L_{12} = \sum_{\bp} v_{\bp}^{2}  \int_{- \infty}^{\infty} d \om 
(- \frac{\partial f}{\partial \om}) \ \om \ A^2\left(\bp, \om \right),
\]
\[
L_{11} = \sum_{\bp} v_{\bp}^{2}  \int_{- \infty}^{\infty} d \om 
(- \frac{\partial f}{\partial \om}) \ A^2\left(\bp, \om \right).
\]
Here ${\bf v}_{\bp} = \partial \e_p/\partial \bp$ is the quasiparticle 
velocity, and $A\left(\bp, \om \right)$ is the spectral function defined as
\[
A\left( \bp, \om \right) = \frac{\t(\om)^{-1}}{\left( \frac{\om}{Z}
- \e_p \right)^2 + \t(\om)^{-2} }.
\]
The evaluation of $L_{11}$ is more straightforward and we will examine it 
first. The momentum sum can be converted into an integral over various energy 
surfaces. The dominant contribution is from the Fermi level, and we get
\[
L_{11} = v_F^2 \dos(0) \int_{- \infty}^{\infty} d \om (- \frac{\partial f}
{\partial \om}) \t(\om).
\]
We have already noted that over the hot region $\t(\om) \propto \om^{-1}$. For
the frequency integral since $\om \sim T$, we get
\ben
L_{11} \propto \left( \frac{\e_F \om_s}{g_0^2} \right) \frac{v_F^2 \dos(0)}
{T}.
\een
This result~\cite{loh,ros97} simply reiterates what we had noted before, 
that when the hot 
carriers dominate transport, $\Delta \s \propto 1/T$. Now for $L_{12}$, we 
first notice that the expression is odd in frequency. This is because $L_{12}$
is a measure of particle-hole asymmetry in the system. In our calculation we 
will consider as phenomenological input two different sources of such 
asymmetry. One such source is from the density of 
states, so that $\dos(\om) = \dos(0) + \om \dos^{\pm}(0) + O(\om^2/\e_F^3)$,
where $\dos^{\pm}(0) \neq 0$ only if there is particle-hole asymmetry in the 
bare non-interacting system of electrons. The 
second source of asymmetry will be from the quasiparticle lifetime which, 
for the
hot carriers, we write as $\t^{-1}(\om) = (g_0^2/\e_F \om_s) \left| \om \right
|(1+ \t \om)$. Here the second term is a possible particle-hole asymmetric 
term in scattering lifetime. $\t$ is a typical scattering time, and $\om < 
\t^{-1}$. After the energy integral around the Fermi surface we get,
\bea
L_{12} &=& v_F^2 \int_{- \infty}^{\infty} d \om (- \frac{\partial f}
{\partial \om}) \om \t(\om) \dos(\om/Z) \nonumber \\
&=& \left( \frac{\e_F \om_s}{g_0^2} \right) v_F^2 \left\{ T \dos^{\pm}(0)/Z
+ T \dos(0) \t \right\}.
\eea

The first term in the above equation is from the asymmetry in density of 
states, and the second term is from the asymmetry in quasiparticle lifetime. 
We note that
the factor of $1/Z$, which leads to entropy enhancement, is associated with 
the asymmetry in density of states. Thus, the first term is the dominant one
and eventually gives anomalous temperature dependence to thermopower. For this
leading term we can write
\ben
S_t \propto \frac{1}{e} \left( \frac{g_0^2\dos^{\pm}(0)}{\e_F \om_s\dos(0)} 
\right) T \ln (\om_s/\d).
\een
In the regime where $T > \Gamma( p - p_c)$, $S_t/T \propto \ln(1/T)$, as has 
been observed~\cite{benz,pflei} in thermopower measurement on 
$ {\rm CeCu}_{6-x}{\rm Au}_x $.

\begin{figure}[tbp]
\includegraphics[scale=.8]{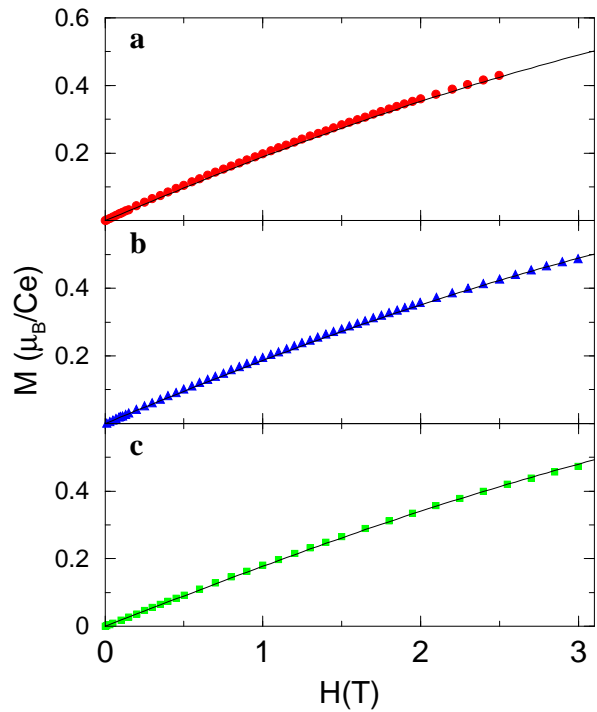}
\caption{Magnetization (M) vs external magnetic field (H) at different
temperatures : a) T=0.15K, b) T=0.3K and c) T=0.8K. The discrete points
are experimental. The solid lines are fits using equation (10).}
\end{figure}

\section{Conclusion}

\begin{figure}[tbp]
\includegraphics[scale=.8]{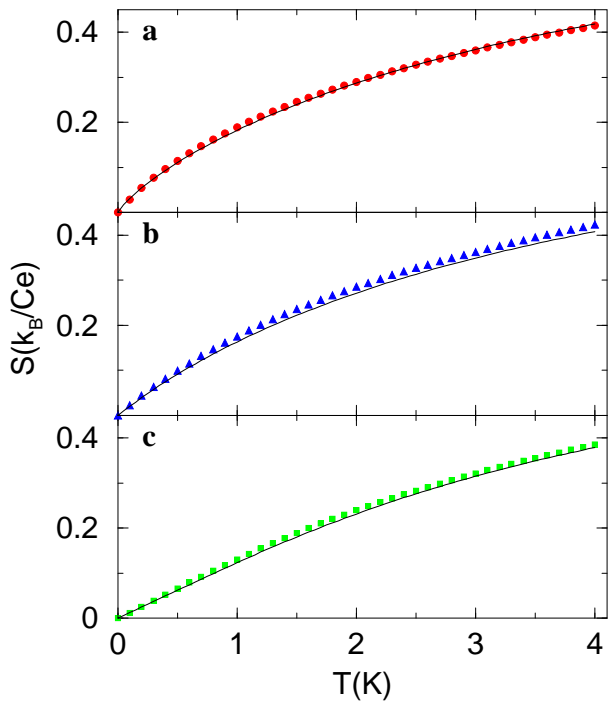}
\caption{Entropy (S) per Ce atom vs temperature (T) at different
magnetic fields : a) H=0T, b) H=1.5T and c) H=3T. The discrete points
are experimental, obtained by numerically integrating data from specific 
heat measurement. The solid lines are fits using equation (10).}
\end{figure}

To check the consistency of our model and calculation, we need  to estimate 
the high energy scale (namely, $T_K^*$) of $ {\rm CeCu}_{5.9}{\rm Au}_{0.1}$.
For this purpose, 
we have  fitted  an approximate form of the free energy function $(F)$
that will
match with the experimental results at low temperature and in the presence of 
magnetic field ($H$). The function that matches well with the experiment has 
the form,
\ben
F(T,H)/k_B = X(T,H) \ln \left[ 2 \cosh \left( \frac{\mu \lambda H}{Y(T,H)} 
\right) \right],
\een
where
\bea
X(T,H) &=& T_{K}^* + C_1 \left( \frac{T^2}{T_{K}^*}\right) 
\nonumber \\ &&
- C_2  \left( \frac{T^2}{T_{K}^*}\right) \ln ( T^2 + C_3 H^2), 
\nonumber \\
Y(T,H) &=& T_{K}^* + (T^2 + C_3 H^2)^{1/2}.
\nonumber 
\eea
Here $C_1$ - $C_3$ are parameters of the fitting function,  $\mu$ is the 
effective magnetic moment of the Ce$^{3+}$ ions in units of Bohr magneton 
($\mu_B$) and $\lambda=\mu_B/k_B =0.67 $. 
We have chosen a simple possible form of the free energy which, at low 
temperatures
($T \ll T_K^*$), is consistent with the critical form of free energy 
that is suggested by renormalization group calculation for 2d spin 
fluctuations,~\cite{millis93} 
namely $F \propto T^2 \ln (T_0/T)$. At high temperatures
($T \gg T_K$) it matches smoothly to an impurity model where the $4f$ cerium 
electrons act as Kondo impurities. The uniform magnetic susceptibility in 
this 
regime is Curie-Weiss like, with $\chi(T) \propto \mu^2/T$.
This temperature dependence is cut-off at $T_K^*$, below which $\chi \sim 
\mu^2/T_K^*$, down to zero temperature. The fitting function is chosen such 
that at very low temperature ($ T \rightarrow 0$), $\chi(T) - \chi(0) 
\propto -T$.~\cite{ioffe} This limiting behaviour agrees with the form 
$\chi \approx a_0 + 1/(a_1 + a_2 T)$ which Rosch {\sl et. al.}~\cite{ros97} 
used to fit susceptibility data up to $1.4$ K. We also find that 
susceptibility derived from equation (10) can describe reasonably well 
(with a difference of at most twenty percent) the 
data~\cite{schr98} up to $6$ K.  
The variation of entropy and magnetization as functions of 
temperature and 
magnetic field that one expects from the above free energy matches well 
with the experiments (Figures 1,2). From the fit we estimate $T_K^*$ to be
around $15$ K, and $\mu \sim 2.6$. 
In the absence of magnetic field the specific heat coefficient ($\g = C/T$)
can be written as $\g = a \ln(T_0/T)$. From the fit we estimate $a=0.5$ J/mol-K$^2$ and $T_0 =9.4$ K, which have comparable order of magnitudes with the 
experimentally measured values $a=0.6$ J/mol-K$^2$ and $T_0 =5.3$ K.~\cite
{loh} The logarithmic behaviour in specific heat
and thermopower in $ {\rm CeCu}_{6-x}{\rm Au}_x $ is observed around 1 K, 
which
is well below $T_K^*$. The experimental fits and the estimates suggest that 
the spin-fermion model that we have been considering is consistent with the 
experimental data.

We now discuss the limitations of our calculation.
We have completely ignored the interaction between the 
spin fluctuations (the $u_0$ term). This is justified
since this  term is marginally irrelevant. 
In our calculation we considered  only the lowest order diagram
in the perturbation series in terms of the spin-fermion coupling. However,
we have examined the lowest order spin-fermion vertex correction, and found 
that it is well-behaved close to the QCP (described in the appendix).  So we 
believe that the qualitative 
features of our calculation will not be modified by including higher order
terms of the series. This is very different from what is found in
the 2d-spin-2d-fermion model, where the spin-fermion vertex
is singular indicating a potential breakdown of the approach.\cite{chubu1} 
So, if the  2d-spin  3d-fermion model
breaks down, there is no trace of this breakdown
in perturbation theory.

From our calculation we see that irrespective of whether the system is clean 
or dirty, if there is a large enough hot region in the system, then both 
specific heat
and thermopower should show anomalous logarithmic temperature dependence. 

Since the microscopic origin of the 2d  spin 
fluctuations is not known, our model seem to be a fine tuned one rather than 
one
that is expected intuitively. It would be interesting to investigate the 
origin of the 2d magnetic coupling, and why most of the Fermi surface is hot
by means of microscopic first principles calculations. This study
should be supplemented by an investigation of the 2d-3d dimensional crossover,
to estimate the energy scale at which it is expected to occur. We notice that 
specific heat and resistivity measurements on 
${\rm YbRh}_2{\rm Si}_2$~\cite{trov} seem to 
indicate that the model, with most of the Fermi surface hot,
is quite valid for it. From this we can conclude that we expect 
to see the behaviour $S_t/T \propto \ln(1/T)$ from thermopower measurement on 
${\rm YbRh}_2{\rm Si}_2$, probably over a wider range of temperatures than 
the Ce-material. 

\section{acknowledgements} 

We are very happy to thank A. Rosch for
many useful insights and  suggestions.
We thank A. Schroeder, G. Aeppli, C. Pfleiderer
and A. Rosch for files of experimental data
on $ {\rm CeCu}_{6-x}{\rm Au}_x $, and O. Trovarelli for a useful discussion 
about the structure of  ${\rm YbRh}_2{\rm Si}_2$.
This research was supported by the Petroleum Research
Fund of the American Chemical Society, under  grant 
ACS-PRF$\#33495$-AC5, and by the Division of Materials
Research of the National Science Foundation under
grant DMR-0096462.

\appendix
\section{Spin-Fermion Vertex}

Here we describe the calculation of the spin-fermion vertex and show that at
the QCP ($\d \rightarrow 0$) the vertex is not singular. This is important
because otherwise our perturbative calculation will break down at low 
temperature near the QCP. With a singular vertex, the coupling constant 
between the electrons and the spin fluctuations will get strongly renormalized
at low energy. The qualitative features of the theory will change, in 
particular the electron self energy. We will express the lowest order 
correction to the bare spin-fermion coupling as $g = g_o(1 + \Gamma) $. Since 
we are interested only in the hot electrons and their low energy interaction
with the spin fluctuations, we will calculate the vertex $\Gamma$ with all 
external frequency zero. The expression for the vertex will then be,
\bea
\Gamma = &&
i g_0^2 \sum_{\bk} \int_{-\infty}^{\infty} \frac{d \om}{2 \pi} 
G \left( \bp_1 + \bk, \om \right) 
\nonumber \\
&&
G \left( \bp_2 + \bk, \om \right)
\chi \left( \bQ + a{\bf\hat{x}} + \bk, \om \right).
\eea
Here $\bp_1$ and $\bp_2$ are two hot points that are connected by wave-vector
$\bQ + a{\bf\hat{x}}$. Expressing the linearized spectrum near the two hot
points as $\e_{1\bk}$ and $\e_{2\bk}$, we can rewrite the above expression as
\begin{widetext}
\[
\Gamma = 
4 g_0^2 \sum_{\bk} \int_{0}^{\infty} \frac{d \om}{\pi}
\frac{\om^2}{ \left( \g^2_{\bQ + \bk} + \om^2 \right) \left(  \e_{1\bk}
+ \e_{2\bk} \right) \left( \om + \e_{1\bk} \right) \left( \om + \e_{2\bk} 
\right)}. 
\]
\end{widetext}
It is easy to check by simple dimensional analysis that as $\d \rightarrow 0$,
the above expression is finite. As an estimate we get $\Gamma \propto 
g_0^2 \Lambda^{1/2}/( \e_F^{3/2} \om_s^{1/2})$, where $\Lambda$ is a 
dimensionless cutoff in the momentum space.

\end{document}